\documentclass[cite,figures]{epl}

\title{Overcharging: The crucial role of excluded volume}
\shorttitle{Overcharging: The crucial role of excluded volume}
\usepackage{graphicx}% Include figure file

\author{R. Messina\inst{1}\thanks{E-mail: \email{messina@mpip-mainz.mpg.de}}
        \and E. Gonz\'{a}lez-Tovar\inst{2,3,4}\thanks{E-mail: \email{henry@dec1.ifisica.uaslp.mx}} 
        \and M. Lozada-Cassou \inst{2,4}\thanks{E-mail: \email{marcelo@www.imp.mx}}  
        \and C. Holm \inst{1} \thanks{E-mail: \email{holm@mpip-mainz.mpg.de}}
      }
\shortauthor{R. Messiina \etal}
\institute{
  \inst{1}  Max-Planck-Institut f\"{u}r Polymerforschung - 
            Ackermannweg 10, 55128 Mainz, Germany \\
  \inst{2}  Departamento de F\'{\i}sica, Universidad Aut\'onoma
            Metropolitana-Iztapalapa - 
            Apartado Postal 55-534, 09340 D.F., M\'exico \\
  \inst{3}  Instituto de F\'{\i}sica, Universidad Aut\'onoma de San Luis
            Potos\'{\i} - 
            \'Alvara Obreg\'on 64, 78000 San Luis Potos\'{\i}, M\'exico \\
  \inst{4}  Programa de Ingenier\'{\i}a Molecular, Instituto Mexicano del
            Petr\'oleo - 
            L\'azaro Cardenas 152, 07730 D. F., M\'exico \\
}
\begin{document}
\maketitle

%\date{\today{}}

\begin{abstract}
  In this Letter we investigate the mechanism for overcharging of a single
  spherical colloid in the presence of aqueous salts within the framework of
  the primitive model by molecular dynamics (MD) simulations as well as
  integral-equation theory. We find that the occurrence and strength of
  overcharging strongly depends on the salt-ion size, and the available volume
  in the fluid. To understand the role of the excluded volume of the
  microions, we first consider an uncharged system.  For a fixed bulk
  concentration we find that upon increasing the fluid particle size one
  strongly increases the local concentration nearby the colloidal surface and
  that the particles become $laterally$ ordered.  For a charged system the
  first surface layer is built up predominantly by strongly correlated
  $counterions$. We argue that this a key mechanism to produce overcharging
  with a low electrostatic coupling, and as a more practical consequence, to
  account for charge inversion with \textit{monovalent} aqueous salt ions.
\end{abstract}
%

% other packs , 61.20.Qg, 

%
%
%HWG with the the MAIN TEXT
%
%%%%% 
Overcharging, or charge inversion, is defined as the situation where a charged
colloid (macroion) accumulates close to its surface more counterions than
necessary to compensate its own bare charge. This effect was already
discovered in the beginning of 80's, both by computer simulations
\cite{Megen_Torrie} and analytical studies
\cite{Marcelo_jcp_1982,gonzalestovar85a}.  
Based on \textit{reversed} electrophoretic mobility, some
experimental
\cite{electrophoresis_EXP}
and numerical (MD)
\cite{electrophoresis_MD}
studies provide some hints
for the manifestation of overcharging and its possible experimental relevance. 
More recently, it has regained a considerable attention on the theoretical side
% and has been studied through semiphenomenological theories 
\cite{Shklovskii_PRL_1999a,Shklovskii_PRE_1999b,Nguyen_2000,Marcelo_PRE_RapCom1999,
Nguyen_condmat-0103208,Messina_PRL_2000,Messina_EPJE_2001,tanaka01a,
greberg98a,deserno01b,Marcelo_Macroiones}.

Overcharging is rather well understood for counterions in \textit{salt-free}
or low salt environment, when excluded volume effects play no role. The
underlying physics at zero temperature of such non-neutral systems can be
quantitatively explained with Wigner crystal (WC) theory
\cite{Shklovskii_PRL_1999a,Shklovskii_PRE_1999b,Nguyen_2000,Messina_PRL_2000}.
The basic concept is that the counterions form a two-dimensional lattice on
the macroion surface, and when overcharging counterions are present on this
layer the energy of the system is lowered compared to the neutral case.  This
feature can be directly and \textit{exactly} computed for a small number of
counterions at zero temperature, and was illustrated in Refs.
\cite{Messina_PRL_2000}.  This WC approach remains
qualitatively correct for finite temperature as long as the Coulomb coupling
is very high. The crystal then melts into a strongly correlated liquid,
where the local order is still strong enough to lower the free energy for the
overcharged state.

The situation becomes much more complicated for aqueous systems, where the
coupling is weak and in addition salt is present.  One-particle inhomogeneous
integral equation theories can describe some of these situations fairly well
\cite{Marcelo_jcp_1982,gonzalestovar85a,greberg98a,Marcelo_Macroiones} but the
computed correlation functions do not necessarily give direct insight into the
physical mechanisms behind these effects.
The presence of excluded volume interactions can lead to layering effects
near the macroion, which are known from simple fluid theories. Here, due to
the presence of charge carriers of both signs, this can even lead to layers of
oscillating charge inversions \cite{gonzalestovar85a,greberg98a,deserno00a}.
This overcharging was also observed (by integral-equation and simulation 
\cite{gonzalestovar85a,deserno01b}) for \textit{monovalent} salt-ions of large size. 
However, until today, the basic mechanism of charge inversion for 
dense salt solutions remains unclear.

The goal of this Letter is (i) to study in detail the role of the excluded
volume contribution for the overcharging of a colloidal macroion in the
presence of salts and (ii) to provide a qualitative insight into the mechanism
behind these effects. We find that in the presence of salt the contribution of
excluded volume can be so important that the size of the small ions dominates
the occurrence of overcharging and the overcharging strength increases with
increasing ionic size when the electrolyte concentration is fixed. Even for
monovalent ions with high enough ionic size, we observe overcharging, which
can not be explained with a salt-free WC picture due to the low Coulomb
coupling strength. In order to have the simplest system we study only the
cases where the coion and the counterions have the same size. This will reduce
the effects of depletion forces which lead to nontrivial features already in
neutral hard sphere fluids. 

Our proposed mechanism will rest on the following arguments:  For a fixed salt
concentration, the available volume in the fluid is function of the electrolyte
particle size. More precisely, the entropy of the solution is decreased by
enlarging the size of the salt-ions \cite{Note_entropy}, which enhances
inter-particle correlations. On the other hand the interface provided by the
macroion leads to an increase of microion density close to the macroion, and
promotes there also $lateral$ ordering, even in the absence of strong
electrostatic coupling, similar to a prefreezing phenomenon. However 
entropy alone can never lead to overcharging, since in this
limiting case coions and counterions have the same radial distribution.
But ordering and weak electrostatic correlations can  lead to overcharging, 
as we are going to prove in this Letter. 

In this work we carried out MD simulations and HNC/MSA (hypernetted chain/mean
spherical approximation) integral-equation to study the overcharging in
spherical colloidal systems within the primitive model. In particular, the 
solvent enters the model only by its dielectric constant and its discrete
structure \cite{elshad_2001} is ignored. 
The system is made up of (i) a large macroion with a bare central charge 
$Q=-Z_{m}e$ (with $Z_m>0$) and (ii) symmetric salt ions of diameter $\sigma$
and valence $Z_c$. The system is globally electrically neutral.

For the simulation procedure, all these ions are confined in an impermeable
cell of radius $R$ and the macroion is held fixed at the cell center. In order
to simulate a canonical ensemble, the Langevin thermostat has been used to 
predict the ions trajectory as has been done in 
Refs. \cite{Messina_PRL_2000}.

Excluded volume interactions are taken into account with a purely $repulsive$
(6-12) Lennard-Jones (LJ) potential characterized by the standard $(\sigma,
\epsilon)$ length and energy parameters, respectively, which is cut-off at the
minimum \cite{Note_LJ}. The macroion-counterion distance of closest approach
is defined as $a$.  Energy units in our simulations are fixed by $ \epsilon =
k_{B}T $ (with $T=298$ K).

The pair electrostatic interaction between any pair \textit{ij}, where
$i$ and $j$ denote either a macroion or a microion, reads 
$U_{coul}(r)=k_B T l_{B}\frac{Z_{i}Z_{j}}{r}\: ,$
%%%%%%%%%%%%%%%%%%%%%%%%%%%
%\begin{equation}
%\label{eq.coulomb}
%_{coul}(r)=k_B T l_{B}\frac{Z_{i}Z_{j}}{r}\: ,
%\end{equation}
%%%%%%%%%%%%%%%%%%%%%%%%%%%
where $l_{B}=e^{2}/(4\pi \epsilon _{0}\epsilon _{r}k_{B}T_{0})$ is the Bjerrum
length fixing the length unit. Being interested in aqueous solutions we
choose the relative permittivity $ \epsilon _{r}=78.5 $, corresponding to
$ l_{B}=7.14 $ \AA. 
The time step is $\Delta t=0.01\tau$ with $\tau =\Gamma ^{-1}$, where $\Gamma$ 
is the damping constant of the thermostat.
Typical simulation runs consisted of $2-7\times10^6$ MD
steps after equilibration.

In order to give prominence to the effect of ionic size, we choose to work at
\textit{fixed} distance  of closest approach (between the centers of  macroion
and the salt-ion) 
$a=2l_{B}=14.28$ {\AA}. In this way, the electrostatic correlation induced 
by the colloid remains the same (i.e. fixed macroion electric field at contact) no matter
what the ionic size is. 
%CHANGES
Thus by changing the ionic size $\sigma$ one also changes that colloidal radius
accordingly so that $a$ remains constant.
The salt concentration $\rho$ is given by $N_-/V$
where $V=\frac {4}{3} \pi R^{3}$ is the cell volume and $N_-$ is the number of
coions. We will restrict the present study to $\rho = 1$M salt concentration.
The fluid volume fraction $f$ is defined as
$(N_++N_-)\left(\frac{\sigma}{2R}\right) ^{3}$ where $N_+$ is the number of
counterions.  To avoid size effect induced by the simulation cell we choose a
sufficiently large radius $R=8.2l_{B}$ yielding to more than 1000 mobile
charges. Simulation run parameters are gathered in Table \ref{tab.simu-param}.
The HNC/MSA calculations were performed using the technique presented in
\cite{Marcelo_Macroiones} and references therein. Here it is assumed that the
system size is infinite, and the bulk salt concentration is fixed. In practice
there should be no observable difference in the correlation functions
(between HNC/MSA and MD) close to the macroions, because the wall effects 
die off sufficiently fast.

%%%%%%%%%%%%%%%שש
%TABLE 1
\begin{table}

\caption{Simulation run parameters for the charged fluid ($A-F$) and the neutral
  fluid ($G-H$).}
\label{tab.simu-param}
%\begin{ruledtabular}
\begin{tabular}{ccccc}
\hline 
 parameters   & salt valence&$Z_{m}$& $\sigma /l_{B} $& $f$   \\
\hline 
run \textit{A}&  2:2        & 48    & $ 1 $           & $ 2.3\times 10^{-1} $\\
run \textit{B}&  2:2        & 48    & $ 0.5 $         & $ 2.9\times 10^{-2} $\\
run \textit{C}&  2:2        & 48    &$ 0.25 $         & $ 3.6\times 10^{-3} $\\
run \textit{D}&  1:1        & 10    &$ 1 $            & $ 2.3\times 10^{-1} $\\
run \textit{E}&  1:1        & 48    &$ 1 $            & $ 2.3\times 10^{-1} $\\
run \textit{F}&  1:1        & 48    &$ 0.5 $          & $ 2.9\times 10^{-2} $\\
run \textit{G}& -           & -     & $ 1 $           & $ 2.3\times 10^{-1} $\\
run \textit{H}& -           & -     & $ 0.5$          & $ 2.9\times 10^{-2} $\\
\hline 
\end{tabular}
%\end{ruledtabular}
\end{table}
%%%%%%%%%%%%%%%%%%%%%%%%%%%%%%%%%%%%%ש

%%%%%%%%%%%%%%%%%%%%%%
%FIG 1
\begin{figure}[t]
\onefigure[width = 7.0 cm]{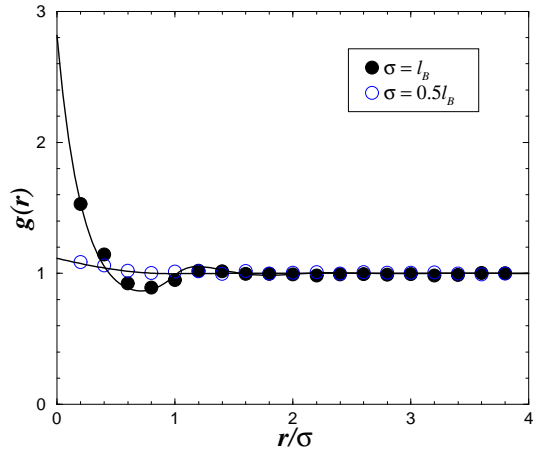}
\caption{Pair distribution function \protect$g(r)\protect$ between 
the colloid and the fluid particle in the neutral state (runs $G-H$) for two particle
sizes $\sigma$.  The origin of the abscissa is taken at the distance of
closest approach \protect$ a=2l_{B}\protect $. Lines and symbols correspond
to HNC theory and simulation respectively.}
\label{fig.gr_HS} 
\end{figure}
%%%%%%%%%%%%%%%%%%%%%
%
We first illustrate the excluded volume correlations present for a $neutral$
hard sphere fluid (runs $G-H$) [identical to systems $A-B$ and $E-F$, but
uncharged (see Table \ref{tab.simu-param})].  Although these kind of systems
are ``simple'' fluids \cite{Hansen_Book_1990}, it is fruitful to elucidate
what exactly happens at this ``low'' fluid density upon varying the fluid
particle size $\sigma$ in the presence of a single large spherical particle.
To characterize the fluid structure we consider the pair distribution function
$ g(r) $ between the colloid and the fluid particle, which is just proportional 
to the local density $n(r)$: $n(r)=\rho g(r)$.  Results are depicted in
Fig.  (\ref{fig.gr_HS}). For the large value $ \sigma= l_B $, one
clearly observes a relative high local concentration as well as a
\textit{short range ordering} nearby the colloidal surface. When the particle
size is reduced by a factor 2 (holding $a$ fixed), the behavior is
qualitatively different and the system is basically uncorrelated. By
increasing $ \sigma $ at fixed fluid density $ \rho $ the mean surface-surface
distance between particles is reduced which in turn leads to higher collision
probability and thus to higher correlations. In other words, by reducing the
\textit{available} volume one promotes \textit{ordering}
\cite{Marcelo_Macroiones,Note_size_asymmetry}.  This seems to be trivial in
the bulk, but the presence of the colloidal surface induces an even enhanced
ordering and the system can prefreeze (order) close to the colloid.
Note that the same effects are
naturally present for a fluid close to a planar wall
\cite{Hansen_Book_1990,Goetzelmann_PRE_1996}.

\begin{figure}[t]
\twoimages[width = 7.0cm]{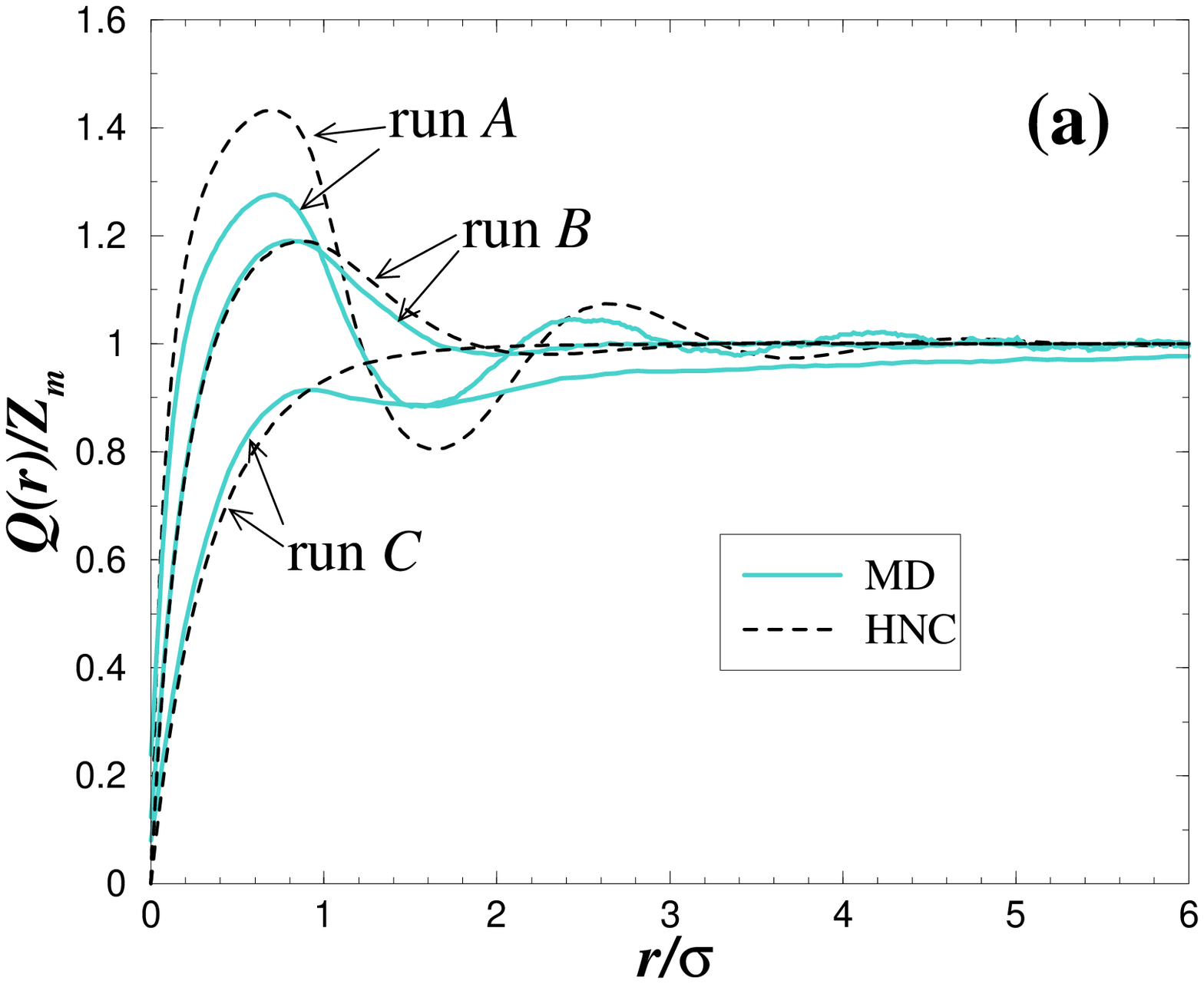}{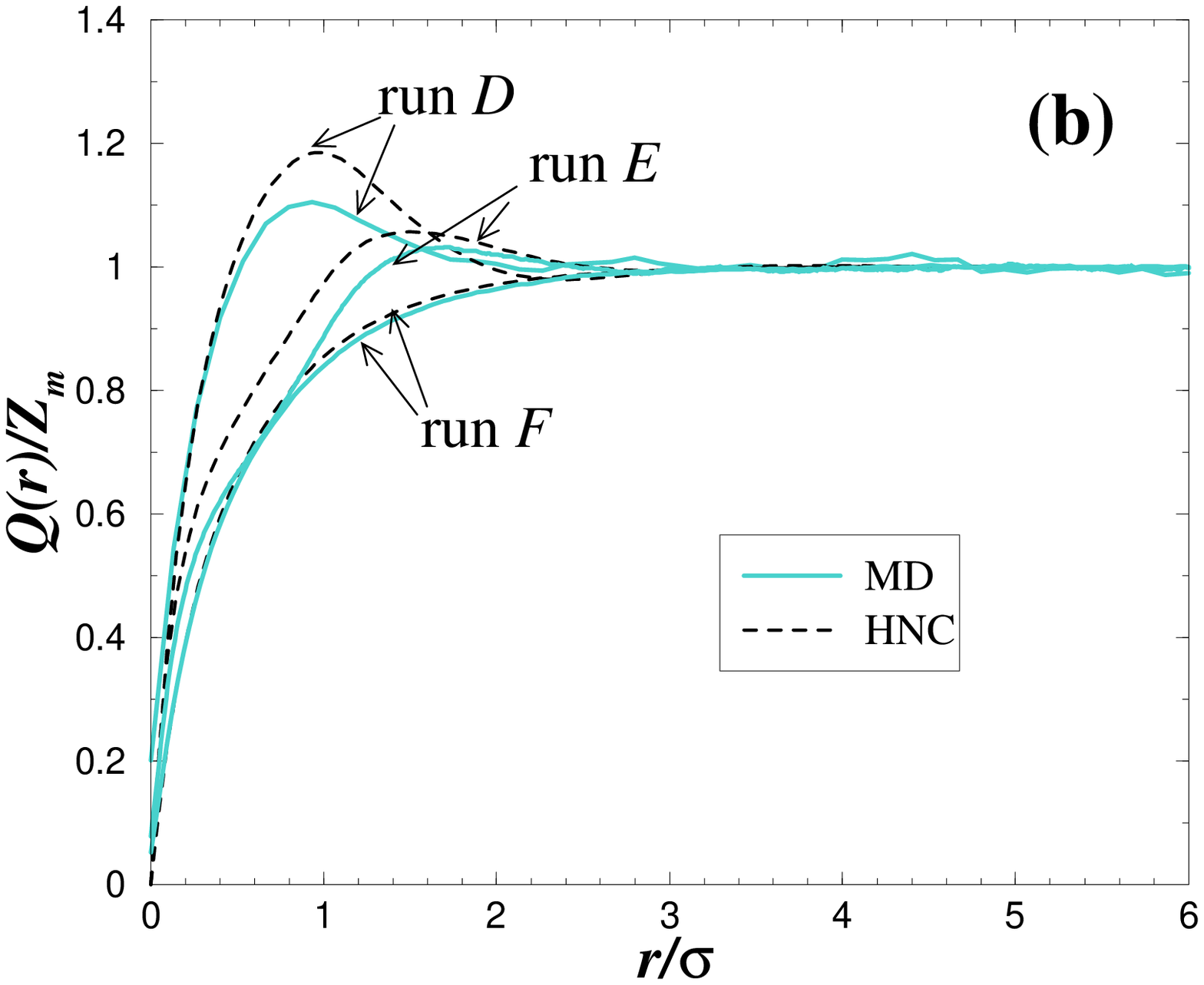}
\caption{
  Reduced fluid integrated charge $Q(r)/Z_{m}\protect$ as a function of
  distance $r$ for for three different particle sizes
  $\sigma$.  (a) divalent salt ions (runs \textit{A-C}), (b) monovalent salt
  ions (runs \textit{D-F}). The origin of the abscissa is taken at the distance
  of closest approach \protect$ a=2l_{B}\protect $}
\label{fig.Qr}
\end{figure}

To characterize the overcharging we introduce the fluid integrated charge $
Q(r) $ which corresponds to the total \textit{net} charge in the fluid
(omitting the macroion bare charge $ Z_{m} $) within a distance $r$ from the
distance of closest approach $ a $. Results are sketched in Fig.
\ref{fig.Qr}(a) and Fig. \ref{fig.Qr}(b) for divalent and monovalent salt ions
respectively.  Theoretical and numerical analysis are in very good qualitative
agreement.

For the divalent electrolyte solutions (runs \textit{A-C}) we observe that
overcharging is strongly dictated by the ionic size $ \sigma $ [see Fig.
\ref{fig.Qr}(a)].  For small ions (run \textit{C}) \textit{no} overcharging
occurs (i. e. $ Q(r)/Z_{m}<1 $), which is a non trivial effect probably
related to the forming of ionic pairs (counterion-coion pairing) due to the
strong pair interaction of $8 k_B T$. This delicate point will be addressed in
a future study, and was also observed in \cite{deserno01b}. Upon increasing $
\sigma $, the degree of overcharging increases. We carefully checked that the
distance $ r =r^{*}$, where $ Q(r^{*}) $ assumes its maximal value, corresponds
to a zone of coion depletion (in average there are less than 2 coions within $
r^{*} $).  This implies that also the absolute number of counterions at the
vicinity of the macroion surface increases with increasing $ \sigma $.  This
is qualitatively in agreement with what we observed above for neutral systems.
However we are going to show later that electrostatic correlations are also
\textit{concomitantly} responsible of this extra counterion population
(increasing with $ \sigma $) in the vicinity of the colloidal surface.

%%%%%%%%%%%%%
%FIG 2
%%%%%%%%%%%%%%%%%%%%%%%%%%%%%%%%

For the monovalent electrolyte solution [see Fig. \ref{fig.Qr}(b)]
overcharging occurs for $ \sigma =l_{B} $ (runs \textit{D} and \textit{E}).
In respect to the salt-free WC picture this is rather unexpected since here
the {}``plasma{}'' parameter $ \Gamma _{cc}=l_{B}Z^{2}_{c}/a_{cc} $, where $
a_{cc}=(\pi c)^{-1/2} $ ($ c $ standing for the two-dimensional surface
counterion concentration) is the mean distance between counterions on the
surface, is small. More precisely for $ Z_{m}=10 $ (run \textit{D}) we find $
\Gamma _{cc}\approx 0.8 $, and for $ Z_{m}=48 $ (run \textit{E}) we have $
\Gamma _{cc} < 1.0 $ \cite{gamma}. But following this salt-free approach, it
is necessary to have at least $ \Gamma _{cc}>2 $ to get overcharge
\cite{rouzina96a}.  
Note that from run $D$ ($Z_m=10$) to run $E$ ($Z_m=48$) one increases the macroion 
surface-charge density leading (for $Z_m=48$) to a higher \textit{absolute} overcharging 
$Q(r^*)$ but a weaker ratio $Q(r^*)/Z_m$, which is qualitatively in accord
with the WC picture, since the maximal overcharging is proportional to $\sqrt{c}$.
A closer look on Fig.  \ref{fig.Qr}(b) reveals that for $
Z_{m}=48 $ (run $E$) $ r^{*}/\sigma $ is shifted to the right compared to the
divalent case. This is merely a packing effect and it is due to the fact that
for monovalent counterions the macroion charge (over)compensation involves
twice more number of counterions particles than in the divalent case.
Therefore for the macroion charge density under consideration ($ Z_{m}=48 $),
more than one counterion-layer is needed to compensate the macroion charge.
Again for a smaller ionic salt size $ \sigma =0.5l_{B} $ (run \textit{F}) the
overcharging effect is canceled.
%

%CHANGES
Recently a depletion of salt-ions (\textit{total} local density of co and
counterions) near the macroion surface was reported for bulk salt-ion
concentrations similar to ours but very \textit{low} surface charge density
\cite{Elshad_EPL}. For all of our investigated cases (runs \textit{A-F}),
however, we never observe such a \textit{depletion}.  The reason is that such
an effect vanishes as soon as the ionic size is sufficiently large and/or the
surface-charge density is large enough, which is in agreement with our
findings.  Indeed, similarly to what happened with image charges, the
self-image repulsion is only relevant for \textit{low} charge density
\cite{Messina_JCP}.  Note that the parameters leading to overcharging {\it
  cannot} give rise to an observation of salt-ion depletion.
 
%%%%%%%%%%%%%%%%%%%%%%%%
% FIG 3
\begin{figure}
\onefigure[width = 7.0 cm]{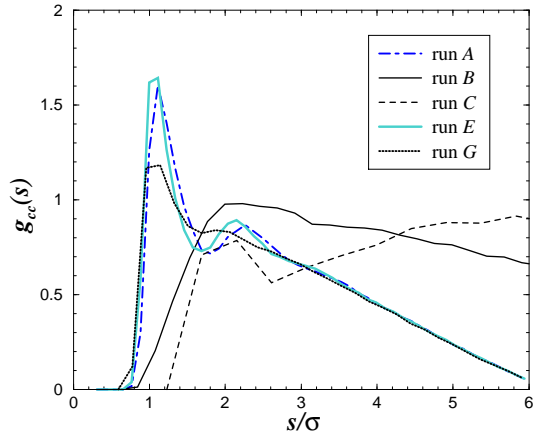}
\caption{Two-dimensional surface counterion correlation functions \protect$g_{cc}(s)\protect $
  for divalent salt (run $A$ - $C$), monovalent salt (run $E$), and a neutral
  system (run $G$).}

\label{fig.gr-CC-2D}
\end{figure}
%%%%%%%%%%%%%%%%%%%%%%%%%

For the WC picture to be effective we need strong lateral correlations which
cannot come from pure electrostatic effects. To see if such correlations are
present we consider the local two-dimensional surface counterion structure.
We analyzed in our simulations the \textit{two-dimensional} counterion pair
distribution $ g_{cc}(s) $, where $ s $ is the arc length on the macroion
sphere of radius $ a $. All counterions lying at a distance $ r<a+0.5\sigma $
from the macroion center are radially projected to the contact sphere of
radius $ a $. Predominantly counterions are present in the first layer.  For
the neutral system (run $G$) we analyzed the small neutral species. Results
are given in Fig. \ref{fig.gr-CC-2D}.  We observe that all systems with an ion
diameter $\sigma=l_B$ show their first peak at roughly $1 \sigma$, and a weaker
second peak at about $2\sigma$, exhibiting long range surface correlations. The
second peak is very weak in the neutral system $G$, but clearly visible in
system $A$ and $E$. Due to the stronger electrostatic repulsion the second
peak for the divalent system $A$ is somewhat further apart than for the
monovalent system $E$. The systems $B$ and $C$ with smaller ion diameters show
a correlation hole of size $\approx 2\sigma$, which is of purely electrostatic
origin.  Therefore lateral correlations can be produced either by pure entropy
effects (run $G$) or pure electrostatic effects (run $C$), or we can have an
enhanced lateral ordering due to the interplay of both (run $A,E$).
Qualitatively, WC arguments are still applicable to monovalent systems (such
as run $E$) if one considers an effective low local surface temperature
stemming from the strong surface ordering.

In summary, the observed overcharging for low Coulomb coupling can be traced
back to the complicated interplay of entropy and energy by the following two
effects. First, by enlarging the excluded volume of the salt ions at fixed
concentration, one lowers the accessible volume to the fluid particles.
Second, the presence of the large macroion provides an interface near which
the density of the fluid is increased compared to the bulk, and the solution
can prefreeze due to entropic effects, provided the available volume gets low
enough.  The closest layer to the interface possesses already strong lateral
correlations, even for a neutral system. If the system is additionally
charged, then even weak Coulomb correlations can lead to the formation of a
strongly correlated liquid, where the overcharged state is energetically
favorable, as shown in
\cite{Shklovskii_PRE_1999b,Messina_PRL_2000,Messina_EPJE_2001,Marcelo_Macroiones}.
The order of this counterion layer is however not created by electrostatic
interactions as in the normal WC picture, but it is largely due to entropic
effects.  The observed overcharging effect might have implications for the
stability of colloidal suspensions. Additions of monovalent salt will
eventually make colloidal suspensions unstable due to the onset of the
van-der-Waals attractions. Upon addition of even more monovalent salt there is
the possibility of salting the precipitate again in, as has been seen for
polyelectrolyte systems\cite{eisenberg59a}. The observation of such a
reentrant transition could be an important hint towards the existence of
overcharging with monovalent salt ions.

\acknowledgments
We are grateful to K. Kremer for a critical reading of the manuscript, and
R. M. thanks the LEA for financial support. 
E. G.-T. acknowledges the support by \textit{PROMEP} and \textit{FAI-UASLP}.
%\bibliographystyle{prsty}
%\bibliography{/people/thnfs/homes/messina/paper6/colloid,/people/thnfs/homes/pep/bibtex/polyelectrolyte}

\begin{thebibliography}{10}
\bibitem{Note_entropy}
This argument holds far enough from the solid-liquid transition as
in the present study (i. e. the highest fluid volume fraction $f$ we
consider is 0.23). Indeed entropy alone can drive crystallization as in
hard-sphere systems.
\bibitem{Note_LJ} 
The LJ potential is rigorously speaking
not a hardcore one, but it can be considered in practice as such.
\bibitem{Note_size_asymmetry}
This feature holds only if the fixed particle (here the colloid) is
larger than the fluid particles.
\bibitem{gamma}
This is due to a packing effect which imposes that $a_{cc}$ cannot be smaller
than $\sigma$.
\end{thebibliography}

%\newpage
\end{document}